\documentclass[conference]{IEEEtran}
\IEEEoverridecommandlockouts
\usepackage{cite}
\usepackage{amsmath,amssymb,amsfonts}
\usepackage{algorithmic}
\usepackage{graphicx}
\usepackage{textcomp}
\usepackage{glossaries}
\usepackage{balance}
\usepackage{siunitx}
\usepackage{booktabs}
\usepackage[caption=false,font=footnotesize]{subfig}

\newacronym{FAS}{FAS}{fluid antenna system}
\newacronym{SNR}{SNR}{signal to noise ratio}
\newacronym{GP}{GP}{Gaussian processes}
\newacronym{BLUE}{BLUE}{best linear unbiased estimate}
\newacronym{i.i.d.}{i.i.d}{independent and identically distributed}
\newacronym{CSI}{CSI}{channel state information}
\newacronym{NMSE}{NMSE}{normalized mean squared error}
\newacronym{RF}{RF}{radio frequency}
\newacronym{pdf}{pdf}{probability density function}
\newacronym{mmWave}{mmWave}{millimeter-wave}
\newacronym{BS}{BS}{base station}
\newacronym{PSA}{PSA}{perturbation sensitivity analysis}
\newacronym{IMPM}{IMPM}{internal multiport method}

\usepackage[dvipsnames]{xcolor}
\def\BibTeX{{\rm B\kern-.05em{\sc i\kern-.025em b}\kern-.08em
    T\kern-.1667em\lower.7ex\hbox{E}\kern-.125emX}}

\begin{document}
\bstctlcite{BSTcontrol}

\title{Radiation-Pattern-Based Channel Estimation for Reconfigurable Pixel Antennas\\

}
\author{\IEEEauthorblockN{Daniel Dinis, Leo M. Lahti, Pasi Ylä-Oijala, Risto Wichman}
\IEEEauthorblockA{
\textit{School of Electrical Engineering, Aalto University}\\
E-mails: \{daniel.1.dinis, leo.m.lahti, pasi.yla-oijala, risto.wichman\}@aalto.fi}
}

\maketitle

\begin{abstract}
Reconfigurable pixel antennas provide spatial diversity through multiple radiation patterns generated by a single antenna aperture, but exploiting this diversity can require substantial pilot overhead for exhaustive channel estimation. This paper jointly considers beam-steerable pixel-antenna design and its communication-side channel estimation. A pixel antenna is designed to generate diverse radiation patterns, whose electromagnetic correlation is then exploited to interpolate unobserved channel responses from a limited number of measurements. The proposed approach is evaluated in terms of \gls{SNR} gain and \gls{NMSE}, demonstrating that effective pattern selection can be achieved with significantly reduced channel-observation overhead and can even outperform exhaustive scanning at low \gls{SNR}.
\end{abstract}


\section{Introduction}

Reconfigurable antennas have attracted significant interest as a means of providing additional spatial diversity without requiring multiple \gls{RF} chains\cite{Wong2021,WongTutorial2025}. Among these, pixel antennas employ a dynamically reconfigurable radiating structure in which the electromagnetic properties of an antenna are tuned using electrically controlled switches. By changing the switch configuration, the antenna can alter its beam direction and radiation characteristics, providing a set of candidate radiation patterns that can be exploited for wireless communication\cite{Zhang2025_pixel,Liu2025_MetaFluid,Rodrigo2016pixel}.

Recent advances in pixel-antenna design have demonstrated the ability to generate diverse and steerable radiation patterns from a single antenna structure. In particular, beam steering enables different pixel configurations to illuminate different spatial directions, potentially providing substantial selection diversity\cite{Antenna_Coding_Shanpu_2026,Liu2026}. However, from a communication-system perspective, exploiting this diversity requires determining which radiation pattern provides the best instantaneous channel response\cite{New2025oversampling}. When a large number of antenna configurations are available, measuring the channel corresponding to every configuration can introduce considerable pilot overhead and channel-estimation latency.

Unlike conventional antenna arrays, where multiple antenna elements can be connected to dedicated \gls{RF} chains and their channel coefficients estimated simultaneously, a reconfigurable pixel antenna typically uses a single reconfigurable aperture and \gls{RF} chain. Its candidate radiation patterns must therefore be evaluated sequentially during pattern selection. As the number of available configurations increases, exhaustive scanning becomes increasingly inefficient despite the relatively small physical size of the antenna.

This challenge of reducing the channel-observation overhead has also been investigated in the context of \gls{FAS}. Several works have considered efficient channel estimation and antenna selection for FAS, including systems based on mechanically movable antennas, liquid-based antennas, and other forms of reconfigurable radiating elements \cite{Zhu2024mov,Bakar2021,Ding25pinching}. These approaches exploit spatial diversity by sampling the channel at a subset of possible antenna positions and using the resulting observations to identify a favorable position \cite{Chai2022b,Zhang2023,Zhang2024Bayesian,Dinis2026_time}. Such spatial approaches typically rely on channel correlation models such as the Jakes model\cite{Jakes1994}, which are not directly suited to pixel antennas, where channel variations arise from changes in the electromagnetic configuration of a fixed aperture rather than physical displacement. Although pixel antennas share the objective of exploiting reconfigurability for selection diversity, their underlying mechanism is fundamentally different. The channel-estimation problem for pixel antennas warrants a dedicated investigation, particularly regarding how the electromagnetic diversity and correlation between radiation patterns can be exploited to reduce the required number of channel observations.

In this work, we jointly investigate a beam steering reconfigurable antenna design problem and the channel-estimation problem arising from the antenna's multiple radiation patterns.
We design a beam-steerable pixel antenna to work at \SI{7.8}{\giga\hertz} using the modified procedure from~\cite{EUCAP2026} with novel beam shaping objective functions. The resulting radiation-pattern diversity is exploited to reduce the channel-observation overhead required for pattern selection. Rather than measuring the channel response of every available radiation pattern, only a subset of patterns is observed, while the remaining channel responses are estimated through interpolation based on the correlation between their radiation patterns. The effectiveness of the proposed approach is evaluated in terms of the resulting \gls{SNR} gains and \gls{NMSE}. The results show that accurate pattern selection can be achieved with limited channel observations and, in some cases, the interpolation-based approach can even outperform a full pattern scan, demonstrating the potential of exploiting radiation-pattern diversity for efficient channel estimation.


\section{System Model}\label{sec:system}

We consider a downlink point-to-point system in which a \gls{BS} communicates with a receiver equipped with a reconfigurable pixel antenna. The \gls{BS} is assumed to have a single transmit antenna, while the receiver can operate in one of multiple radiation configurations obtained by changing the states of the antenna switches. Each configuration produces a distinct radiation pattern and therefore corresponds to an effective communication channel. 
The channel observed by the receiver can then be represented by the vector
\begin{equation}
\mathbf{h} = [h_1,h_2,\ldots,h_N]^\mathsf{T} \in \mathbb{C}^{N},
\end{equation}
where $h_n$ denotes the complex channel coefficient corresponding to the $n$th radiation pattern. 

We assume a narrowband, static channel during each channel-estimation interval. The channel coefficients are generated from the electromagnetic radiation patterns of the pixel antenna and a random incident field. 
A random incident field is represented by the complex coefficient vectors $\mathbf{s}_{\varphi}$ and $\mathbf{s}_{\theta}$, which are modeled as independent zero-mean complex Gaussian variables. The resulting channel coefficients can therefore be expressed as
\begin{equation}
\mathbf{h}
=
\mathbf{s}_{\varphi}^\mathsf{T}\mathbf{E}_{\varphi}
+
\mathbf{s}_{\theta}^\mathsf{T}\mathbf{E}_{\theta},
\end{equation}
where the columns of $\mathbf{E}_{\varphi}$ and $\mathbf{E}_{\theta}$ contain the corresponding discrete electric field components of the available radiation patterns. The generation of these field vectors from the antenna configurations is described in Section~\ref{sec:design}.

During channel estimation, the receiver observes a noisy version of the channel corresponding to the selected radiation configurations. For a transmitted pilot symbol $x$, the observation associated with pattern $n$ is
\begin{equation}
y_n = h_n x+w_n,
\end{equation}
where $w_n\sim\mathcal{CN}(0,\sigma_w^2)$ denotes additive white Gaussian noise. 

The different switch configurations of the pixel antenna produce distinct electromagnetic radiation patterns. Since these configurations do not correspond to different physical antenna positions, their channel correlation cannot be characterized through spatial separation. Instead, we quantify the similarity between two configurations from their radiated electric fields. Let $\mathbf{e}_i(\theta,\varphi)$ and $\mathbf{e}_j(\theta,\varphi)$ denote the continuous vector electric fields of the $i$th and $j$th radiation patterns, respectively. Their normalized correlation is defined as
\begin{equation}
\hat{\rho}_{ij} = \frac{\iint_{S} |\mathbf{e_i}(\theta, \varphi)\cdot\mathbf{e_j}^*(\theta, \varphi)|\text{d}S}{
\iint_{S} |\mathbf{e_i}(\theta, \varphi)|^2\text{d}S
\iint_S |\mathbf{e_j}(\theta, \varphi)|^2\text{d}S
}.
\end{equation}

The resulting pattern-correlation matrix is given by
\begin{equation}
\mathbf{R}=\left[\hat{\rho}_{ij}\right]_{i,j=1}^{N}
\end{equation}
and is obtained directly from the simulated radiation patterns of the pixel antenna. For any subset of patterns $\mathcal{P}\subseteq\{1,\ldots,N\}$, we denote the
corresponding submatrix of $\mathbf{R}$ by
$\mathbf{R}_{\mathcal{P},\mathcal{P}}$, and the matrix containing the rows indexed
by $\mathcal{P}$ and all columns by $\mathbf{R}_{\mathcal{P},:}$. Thus,
\begin{equation}
\mathbf{R}_{\mathcal{P},\mathcal{P}}
=
\mathbf{R}(\mathcal{P},\mathcal{P}),
\qquad
\mathbf{R}_{\mathcal{P},:}
=
\mathbf{R}(\mathcal{P},:).
\end{equation}

\section{Antenna Design}\label{sec:design}
Reconfigurable antennas such as pixel antennas can be divided into frequency reconfigurable, radiation pattern reconfigurable (beam steering, beam shaping, polarization)
and compound reconfigurability~\cite{Reconfigurable_Bernha_2007}.
Pixel antennas can achieve different forms of reconfigurability by changing the switch configuration~\cite{Rodrigo2016pixel}.
Typically pixel antennas are optimized using methods such as genetic algorithms~\cite{Multiport_Pixel_Jiang_2022}, successive exhaustive
boolean optimization~\cite{Successive_Bool_Shen_2017}, and \gls{PSA}~\cite{Pixel_Antenna_O_Jiang_2022,EUCAP2026}.
In this work we design a radiation pattern reconfigurable (beam-steerable) pixel antenna at \SI{7.8}{\giga\hertz}  with pixel states optimized using
\gls{PSA} based on beam steering and shaping target.

\subsection{Antenna Model}

\begin{figure}[htpb]
\begin{center}
\resizebox{.3\textwidth}{!}{\includegraphics{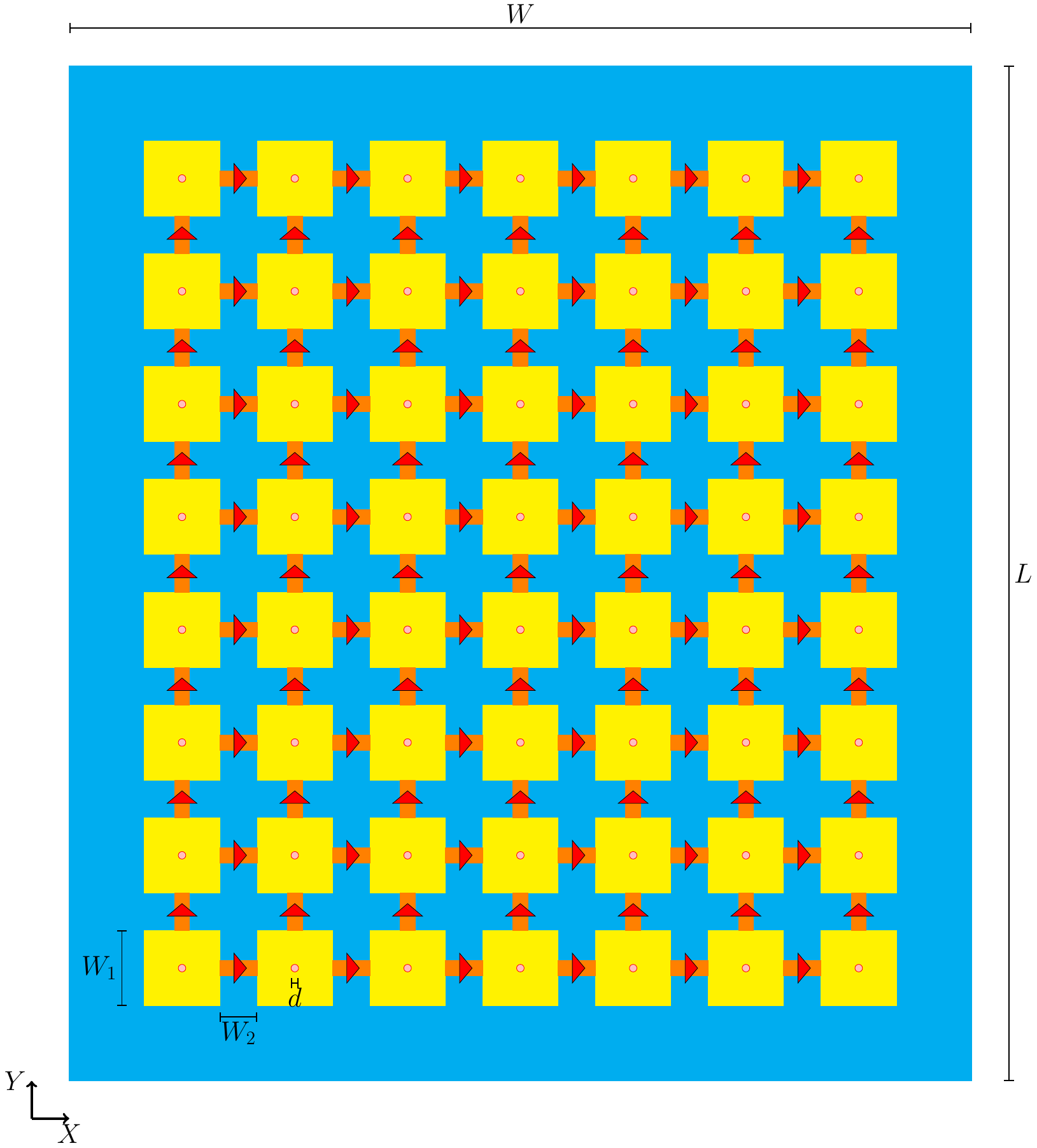}}\\
\end{center}
\caption{Top view of the pixel antenna model. Blue is the substrate Rogers RO4003C of thickness $\tau$, yellow are 0-thickness PEC pixels,
orange are switches, pink are feeds, and red are simulation ports. Adapted from~\cite{EUCAP2026}.}
\label{fig:antmod}
\end{figure}

The base pixel antenna model used to find beam-steerable design is a probe fed, $8 \times 7$ square pixel structure with vertical and horizontal switches.
The top view of the geometry is presented in Fig.~\ref{fig:antmod} and the cross section is visible in Fig. 1 (b) of~\cite{EUCAP2026} with physical dimensions listed in Table~\ref{tab:dims}. 

\begin{table}[htpb]
  \caption{Antenna model dimensions from Fig.~\ref{fig:antmod} and Fig. 1 (b) of~\cite{EUCAP2026}.}
\label{tab:dims}
\centering
\begin{tabular}{|c|c|}
\hline
Dimension & $\si{\milli\meter}$ \\
\hline
$W$ & 42.107\\
\hline
$L$ & 48.308\\
\hline
$W_1$ & 3.77\\
\hline
$W_2$ & 1\\
\hline
$d$ & 1.5\\
\hline
$\tau$ & 3\\
\hline
\end{tabular}
\end{table}
The pixels and ground plane are modeled as 0-thickness perfect electric conductor (PEC), switches as square face ports and feeds as cylindrical face ports.
Every single pixel has potential feed connected from the ground plane to the center of the pixel.
We first select a single feed to be driven and terminate the rest.
There are $N_s = 2\cdot8\cdot7 - 8 - 7 = 97$ switches and $N_f = 8\cdot7 = 56$ feeds, total of $N_p = N_s + N_f = 153$ simulation ports.
The scattering parameters (S-parameters)~\cite{pozar} of the full structure are divided into blocks
\begin{equation}
  \mathbf{S} = \begin{bmatrix}
    S_{dd} & \mathbf{S_{dl}} \\
    \mathbf{S_{ld}} & \mathbf{S_{ll}}
  \end{bmatrix}
  \in \mathrm{C}^{N_p \times N_p}
\end{equation}
where $S_{dd}$ is a complex scalar number representing the reflection coefficient of the driven feed, $\mathbf{S_{ll}}$ describes the behavior of the switches and terminated feeds
while $\mathbf{S_{dl}}$ and $\mathbf{S_{ld}}$ represent the coupling between the driven input, switches, and terminated feeds.

The terminations of both switches and non-driven feeds are done via S-parameter \gls{IMPM}~\cite{Analysis_and_opt_Abardo_2024}
\begin{equation}
\label{eq:IMPM}
S_{\text{IMPM}} = S_{dd} + \mathbf{S_{dl}}\mathbf{\Gamma(x)}(\mathbf{1} - \mathbf{S_{ll}\Gamma(x)})^{-1}\mathbf{S_{ld}}
\end{equation}
where $\mathbf{\Gamma}(\mathbf{x})  \in \mathrm{C}^{N_p - 1 \times N_p - 1}$ is diagonal matrix
of reflection coefficients as a function of binary switch states $\mathbf{x}$
together with open terminations to non-driven feeds 
\begin{equation}
    \mathbf{\Gamma}(\mathbf{x})
    = \text{diag}([
        1 - 2x_1, 1 - 2x_2, \dots, 1 - 2x_{N_s}, 1, \dots, 1]).
\end{equation}
Switches are assumed to be ideal and thus open switches, together with terminated feeds, are terminated with open circuit
reflection coefficient $\Gamma_{ii} = 1$ and closed switches with short circuit reflection coefficient $\Gamma_{ii} = -1$.
From this we can denote the effective feeding coefficients, i.e., coefficients of the incident waves at the terminated ports corresponding to the reflections,
as
\begin{equation}
    \mathbf{a}(\mathbf{x}) = 
    \begin{bmatrix}
        1 \\
        \mathbf{\Gamma}(\mathbf{x})(\mathbf{1} - \mathbf{S_{ll}}\mathbf{\Gamma}(\mathbf{x}))^{-1}\mathbf{S_{ld}}
    \end{bmatrix}.
\end{equation}
This can be used to construct the continuous electric far-field of the $n$th configuration, $\mathbf{e_n}(\theta, \varphi)$, using the 
farfield $\mathbf{e^i}(\theta, \varphi)$ of the $i$th port
\begin{equation}
    \mathbf{e_n}(\theta, \varphi) = [\mathbf{e^1}(\theta, \varphi), \mathbf{e^2}(\theta, \varphi), \dots, \mathbf{e^{N_p}}(\theta, \varphi)]\mathbf{a}(\mathbf{x}).
\end{equation}
We also denote the components of discrete electric fields observed at $N$ directions $(\theta_j, \varphi_j)$ radiated by the
$i$th antenna port as $E_{\theta,j}^i, E_{\varphi,j}^i$ and therefore we can use the matrix
\begin{equation}
    \label{eq:ffmatrix}
    \mathbf{F} =
    \begin{bmatrix}
        E_{\theta,1}^1 & E_{\theta,1}^2 & \dots & E_{\theta,1}^{N_p} \\
        E_{\theta,2}^1 & E_{\theta,2}^2 & \dots & E_{\theta,2}^{N_p} \\
        \vdots         & \vdots &  \ddots & \vdots \\
        E_{\theta,N}^1 & E_{\theta,N}^2 & \dots & E_{\theta,N}^{N_p} \\
        E_{\varphi,1}^1 & E_{\varphi,1}^2 & \dots & E_{\varphi,1}^{N_p} \\
        E_{\varphi,2}^1 & E_{\varphi,2}^2 & \dots & E_{\varphi,2}^{N_p} \\
        \vdots         & \vdots &  \ddots & \vdots \\
        E_{\varphi,N}^1 & E_{\varphi,N}^2 & \dots & E_{\varphi,N}^{N_p}
    \end{bmatrix}
\end{equation}
to map effective feeding coefficients to the observed far-fields 
\begin{equation}
    \mathbf{f} = 
    [\mathbf{E_{\theta,n}},\mathbf{E_{\varphi,n}}]^T
    = \mathbf{F}\mathbf{a}.
\end{equation}

\subsection{Configuration Optimization}

Pixel configurations are optimized using a 2-bit \gls{PSA}~\cite{Pixel_Antenna_O_Jiang_2022}.
This method updates the current configuration as long as its objective function $f(\mathbf{x})$
can be improved by flipping 2 switch states.
Once the current best configuration ca not be improved further, the algorithm terminates.
\gls{PSA} is a greedy deterministic optimization method that has been used for the design of pixel antennas~\cite{Pixel_Antenna_O_Jiang_2022,EUCAP2026}. In this work, we utilize \gls{PSA} in two steps. First, to find a configuration satisfying the matching criterion, and then to utilize that design as an initial configuration for beam steering optimization.

The first objective function is the absolute value of the reflection coefficient~\eqref{eq:IMPM}. Here the initial configuration for \gls{PSA}
is simply all switches open. We run \gls{PSA} to receive a single well matched configuration.
After this we utilize a pattern optimization target
\begin{equation}
    \label{eq:pattarg}
    f_i(\textbf{x}) = \mathbf{a}(\mathbf{x})^H\mathbf{F}^H\mathbf{W^{i}Fa}(\mathbf{x})
\end{equation}
for our target directions.
We introduce a weight matrix $\mathbf{W^i} \in \mathbb{R}^{2N_d\times 2N_d}$, which is a diagonal matrix for desired pattern $i = 1,2, \dots,N$, and
$N_d$ is the number of our observed directions.
The matrix $\mathbf{W^i}$ is built such that each element $W_{jj}^i$ is defined based on the desired radiation direction $i$ 
and the observed direction $(\theta_j, \varphi_j)$
\begin{equation}
    \label{eq:weightmatrix}
    \begin{aligned}
    \mu_j^i &= \cos(\theta_j)\cos(\theta_i) + \sin(\theta_j)\sin(\theta_i)\cos(\varphi_j-\varphi_i) \\
    \gamma_j^i &= \begin{cases}
    \pi & \mu_j^i < -1\\
    \arccos(\mu_j^i)& -1 \leq \mu_j^i \leq 1 \\
    0 & 1 < \mu_j^i \\
\end{cases} \\
    W_{jj}^i &= \begin{cases}
        T\cos^2\frac{\pi\gamma_j^i}{2\gamma_m} & 0 \leq \gamma_i^j \leq \gamma_m \\
        -P\sin^2\frac{\pi(\gamma_j^i - \gamma_m)}{2(\gamma_s - \gamma_m)} & \gamma_m < \gamma_i^j \leq \gamma_s \\
        -P & \gamma_i^j > \max\{\gamma_m, \gamma_s\}
    \end{cases}
    \end{aligned}
\end{equation}
where $T$ is the desired direction weight, $\gamma_m$ is the desired direction angular width, $P$ is the undesired direction penalty, and $\gamma_s$
is the angular width from $W_{jj}^i = 0$ where $W_{jj}^i = -P$.
Therefore we get $N$ different optimization targets, which yield us $N$ different radiation patterns.
We then optimize for every single target function $f_i$ using \gls{PSA}
\begin{equation}
    \begin{aligned}
    &\max f_i(\mathbf{x_i}) \\
    &\text{s.t.} \mathbf{x_i} \in \{0, 1\}^{N_s}
    \end{aligned}
\end{equation}
from which we get our $n$th radiation pattern
\begin{equation}
    \mathbf{f}_n = \mathbf{Fa}(\mathbf{x_n}).
\end{equation}

\section{Channel Estimation}
\label{sec:methods}

The reconfigurable pixel antenna provides a set of distinct radiation patterns that can be exploited for pattern selection. However, measuring the channel response of every available pattern results in a channel-estimation overhead that increases with the number of configurations. To reduce this overhead, we consider measuring the channel response of only $K \ll N$ patterns and estimating the responses of the remaining patterns from these observations using the radiation-pattern correlation matrix $\mathbf{R}$. The estimated responses are then used to select the pattern with the largest channel gain.

\subsection{Selection by Minimizing Interpolation Uncertainty}

The observation set is selected by minimizing the total uncertainty of the interpolated channel responses. A greedy procedure is used, where the set of observed patterns is built sequentially. Let $\mathcal{P}_k$ denote the set of patterns selected after $k$ steps, with $\mathcal{P}_0=\emptyset$. At each step, each pattern not already in $\mathcal{P}_k$ is considered as a candidate. The candidate that results in the lowest total posterior interpolation uncertainty is then added to the observation set.


For a candidate pattern $i\notin\mathcal{P}_k$, define the trial observation set as
\begin{equation}
\mathcal{P}_{k,i} = \mathcal{P}_k \cup \{i\}.
\end{equation}

The corresponding posterior covariance is
\begin{equation}
\mathbf{C}_{\mathrm{post}}^{(i)}
=
\mathbf{R}
-
\mathbf{R}_{\mathcal{P}_{k,i},:}^{\mathsf{H}}
\left(
\mathbf{R}_{\mathcal{P}_{k,i},\mathcal{P}_{k,i}}
+
\sigma_{\mathrm{reg}}^2\mathbf{I}
\right)^{-1}
\mathbf{R}_{\mathcal{P}_{k,i},:},
\end{equation}
where $\sigma_{\mathrm{reg}}^2$ is a small regularization term used for numerical stability. The next pattern is selected as
\begin{equation}
i^\star
=
\underset{i\notin\mathcal{P}_k}{\arg\min}
\;
\operatorname{tr}
\left(
\mathbf{C}_{\mathrm{post}}^{(i)}
\right),
\end{equation}
and the selected set is updated according to
\begin{equation}
\mathcal{P}_{k+1}
=
\mathcal{P}_k\cup\{i^\star\}.
\end{equation}

The procedure is repeated until $K$ patterns have been selected, yielding the final observation set $\mathcal{P}=\mathcal{P}_K$. This set is then used for channel observation and interpolation.

\subsection{Interpolation}

Let $\mathbf{y}_{\mathcal{P}}$ contain the noisy channel observations obtained from the selected patterns. 
Assuming a zero-mean complex Gaussian channel, the channel responses of all available patterns are estimated as
\begin{equation}
\hat{\mathbf{h}}= \mathbf{y}_{\mathcal{P}}^\mathsf{T}
\left( \mathbf{R}_{\mathcal{P},\mathcal{P}} +\sigma_w^2\mathbf{I} \right)^{-1} \mathbf{R}_{\mathcal{P},:}.
\end{equation}

The selected pattern for communication is chosen to maximize the estimated channel gain
\begin{equation}
n^\star = \underset{n\in\{1,\ldots,N\}}{\arg\max}
\;|\hat{h}_n|^2.
\end{equation}

Only the $K$ patterns in $\mathcal{P}$ require channel observations, while the responses of the remaining patterns are inferred through their radiation-pattern correlation. Unlike conventional spatial interpolation, the proposed method does not require antenna positions or a predefined spatial channel covariance model. Instead, the interpolation kernel is obtained directly from the electromagnetic radiation patterns of the pixel antenna.

\section{Performance Results}

\subsection{Antenna performance}
\label{sec:antenna_perf}
Antenna structure presented in Section~\ref{sec:design} was simulated using CST Microwave Studio 2025.
Then the radiation matrix $\mathbf{F}$ and weight matrices $\mathbf{W}^i$ were formed from the directions in Table~\ref{tab:targetssummary} using
equations~\eqref{eq:ffmatrix} and~\eqref{eq:weightmatrix} respectively.
\begin{table}[ht]
  \centering
  \caption{Target directions. First column tells $\theta$ value, second how many targets with given $\theta$, third how much $\varphi$ changes,
  fourth the range of $\varphi$ and last one the indices of targets for given $\theta$.}
  \label{tab:targetssummary}
  \begin{tabular}{| r|r|r|l|r |}
    \hline
    $\theta$ & Targets & $\Delta\varphi$ & $\varphi$ range & Numbering \\  \hline
    $0^\circ$  &  1 & --          & $0^\circ$            & 1 \\ \hline
    $15^\circ$ &  4 & $90^\circ$     & $0^\circ$--$270^\circ$  & 2--5 \\  \hline
    $30^\circ$ &  8 & $45^\circ$     & $0^\circ$--$315^\circ$  & 6--13 \\  \hline
    $45^\circ$ & 16 & $22.5^\circ$   & $0^\circ$--$338^\circ$  & 14--29 \\ \hline
    $60^\circ$ & 32 & $11.25^\circ$  & $0^\circ$--$349^\circ$  & 30--61 \\ \hline
  \end{tabular}
\end{table}
After this \gls{PSA} is run to yield $N = 61$ distinct configurations and radiation patterns. Verification simulations were done using CST such that closed switches were replaced with PEC strips and open switches were removed
together with terminated feeds.

\begin{figure}[!t]
  \centering
  \subfloat[$\theta=0^\circ$]{%
    \includegraphics[width=0.48\columnwidth]{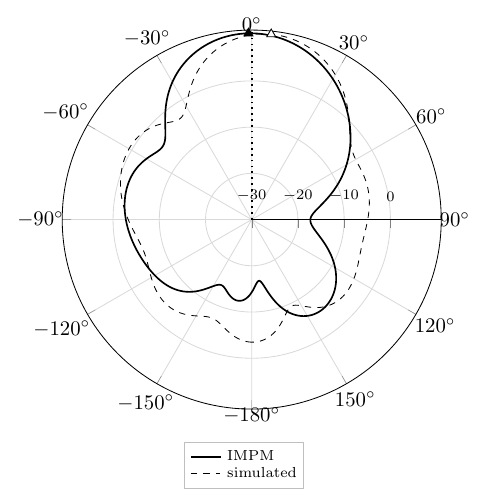}%
    \label{fig:pat0}}
  \\[2pt]
  \subfloat[$\theta=15^\circ$]{%
    \includegraphics[width=0.48\columnwidth]{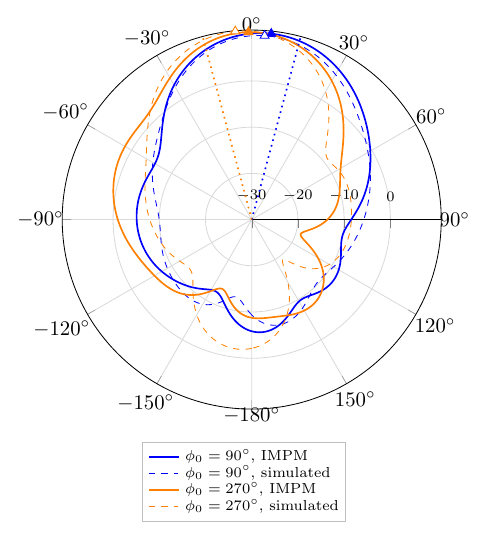}%
    \label{fig:pat15}}
  \hfil
  \subfloat[$\theta=30^\circ$]{%
    \includegraphics[width=0.48\columnwidth]{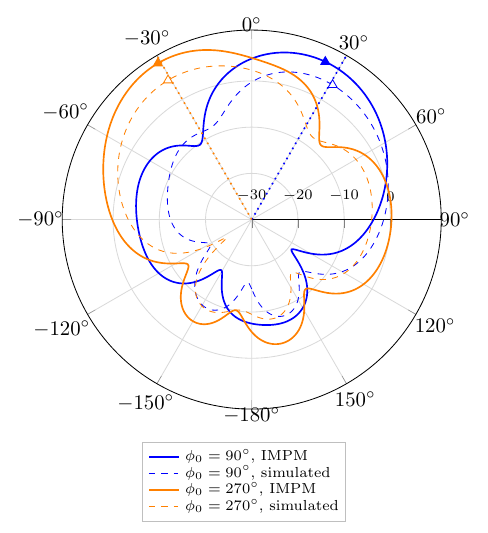}%
    \label{fig:pat30}}
  \\[2pt]
  \subfloat[$\theta=45^\circ$]{%
    \includegraphics[width=0.48\columnwidth]{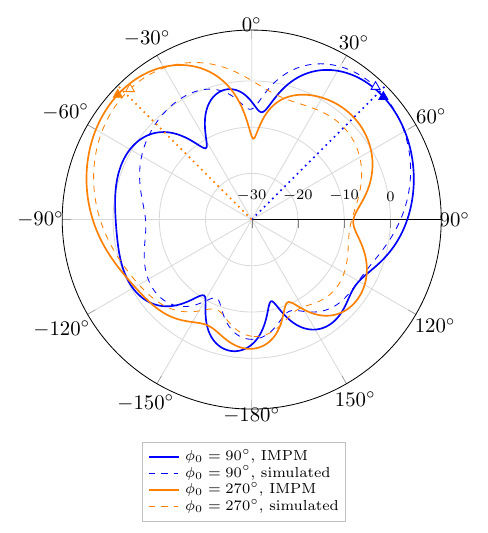}%
    \label{fig:pat45}}
  \hfil
  \subfloat[$\theta=60^\circ$]{%
    \includegraphics[width=0.48\columnwidth]{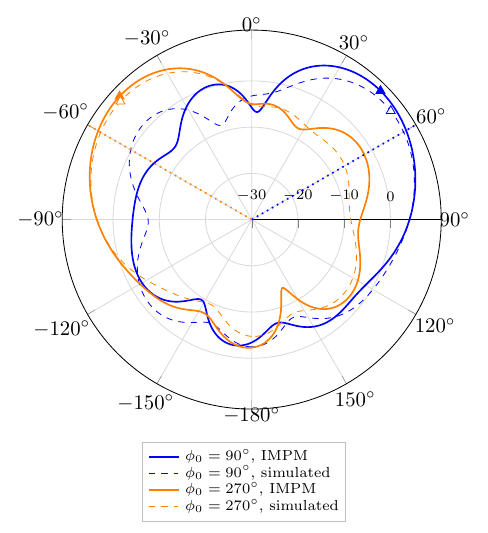}%
    \label{fig:pat60}}
  \caption{YZ-cuts of realized gains of selected target configurations.}
  \label{fig:pats}
\end{figure}

Few selected radiation patterns in the decibel scale are showcased in Fig.~\ref{fig:pats} in YZ-plane
($\varphi$ is either $90^\circ$ or $270^\circ$, $\theta$ is free). The dashed radial lines are the target directions, solid patterns are calculated using \gls{IMPM} and dashed lines show 
simulated verification results.
The simulated $S_{11}$ matching is at best \SI{-22.39}{\decibel}, at worst \SI{-2.12}{\decibel} and has the median value of \SI{-8.72}{\decibel}.

The plotted radiation patterns show good correspondence between \gls{IMPM} and simulated verification results.
They also demonstrate good beam steering capabilities.
While the maxima of the realized gains of an antenna configuration is not necessarily at the desired target, they radiate well enough into the desired directions.

\subsection{Channel estimation performance}

This section evaluates the channel-estimation and pattern-selection performance of the reconfigurable pixel antenna. Unless stated otherwise, the simulations follow the antenna configuration described in Section~\ref{sec:antenna_perf}.

Four pattern-selection schemes are considered:
\begin{enumerate}
    \item \textbf{Proposed interpolation:} $K=10$ radiation patterns are observed during the pilot phase. The observation set is selected by minimizing the total interpolation uncertainty, and the channel responses of all $N$ patterns are subsequently estimated using the radiation-pattern correlation matrix $\mathbf{R}$.

    \item \textbf{Low-correlation selection:} To maximize the amount of information obtained from the measurements, $K=10$ patterns with low mutual radiation-pattern correlation are observed, and the pattern with the largest measured channel gain is selected.

    \item \textbf{Best of all observations:} All $N=61$ radiation patterns are exhaustively scanned using noisy pilot measurements, and the pattern with the largest measured channel gain is selected.

    \item \textbf{Perfect CSI:} The pattern with the largest noise-free channel gain among all $N=61$ patterns is selected. This represents the ideal selection performance and serves as an upper bound.
\end{enumerate}

\begin{figure}[ht]
    \centering
    \includegraphics[width=1.0\linewidth]{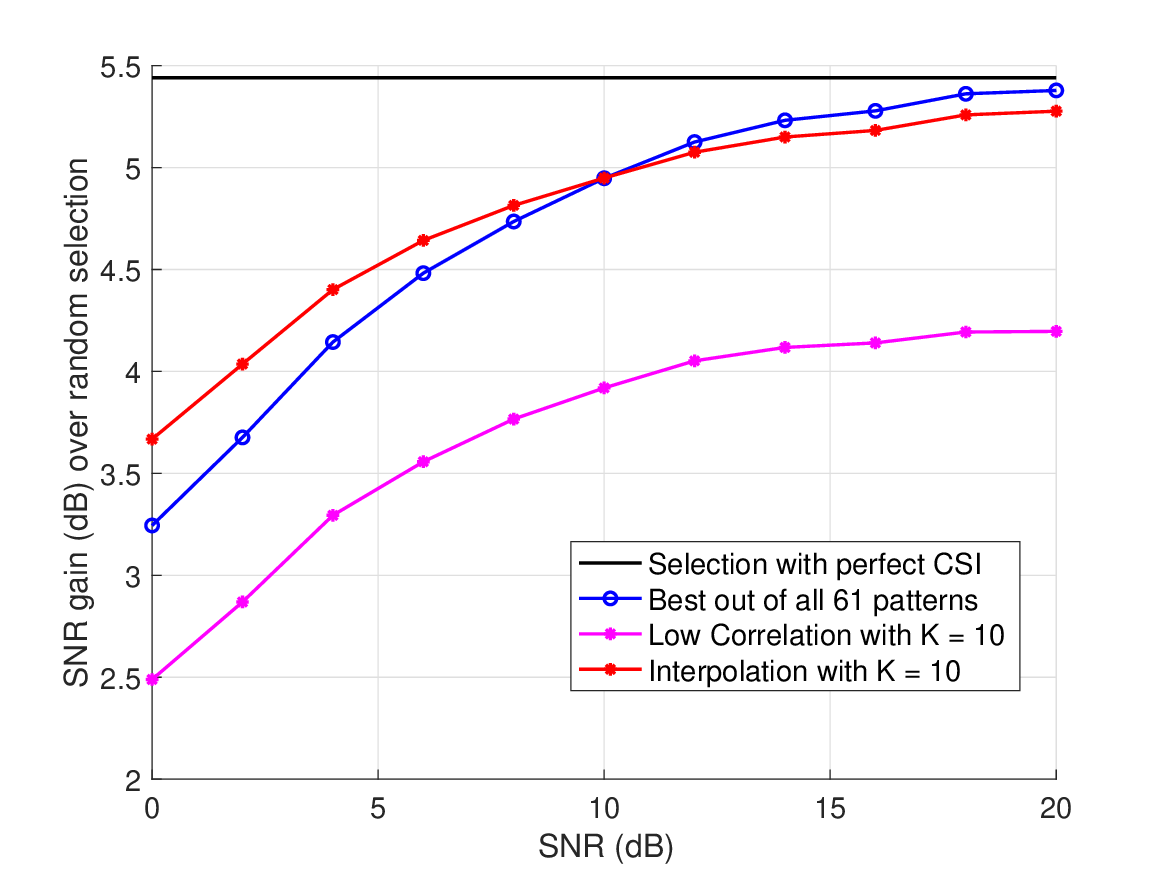}
    \caption{SNR gain as a function of SNR for $K=10$.}
    \label{fig:SNRgain}
\end{figure}

Figure~\ref{fig:SNRgain} shows the average \gls{SNR} gain relative to random pattern selection. The proposed interpolation scheme outperforms direct selection from the low-correlation patterns while using the same number of observations. At low \gls{SNR}, it can even outperform exhaustive scanning, since scanning all $N$ patterns increases the likelihood of selecting a pattern with an artificially high noisy gain. By exploiting the correlation between radiation patterns, interpolation provides more reliable estimates of the unobserved patterns and therefore improves pattern selection. As the \gls{SNR} increases, the noisy channel estimates become more accurate, allowing exhaustive scanning to approach the performance of the perfect-\gls{CSI} benchmark, with both approaches converging toward the ideal selection performance.

\begin{figure}[ht]
    \centering
    \includegraphics[width=1.0\linewidth]{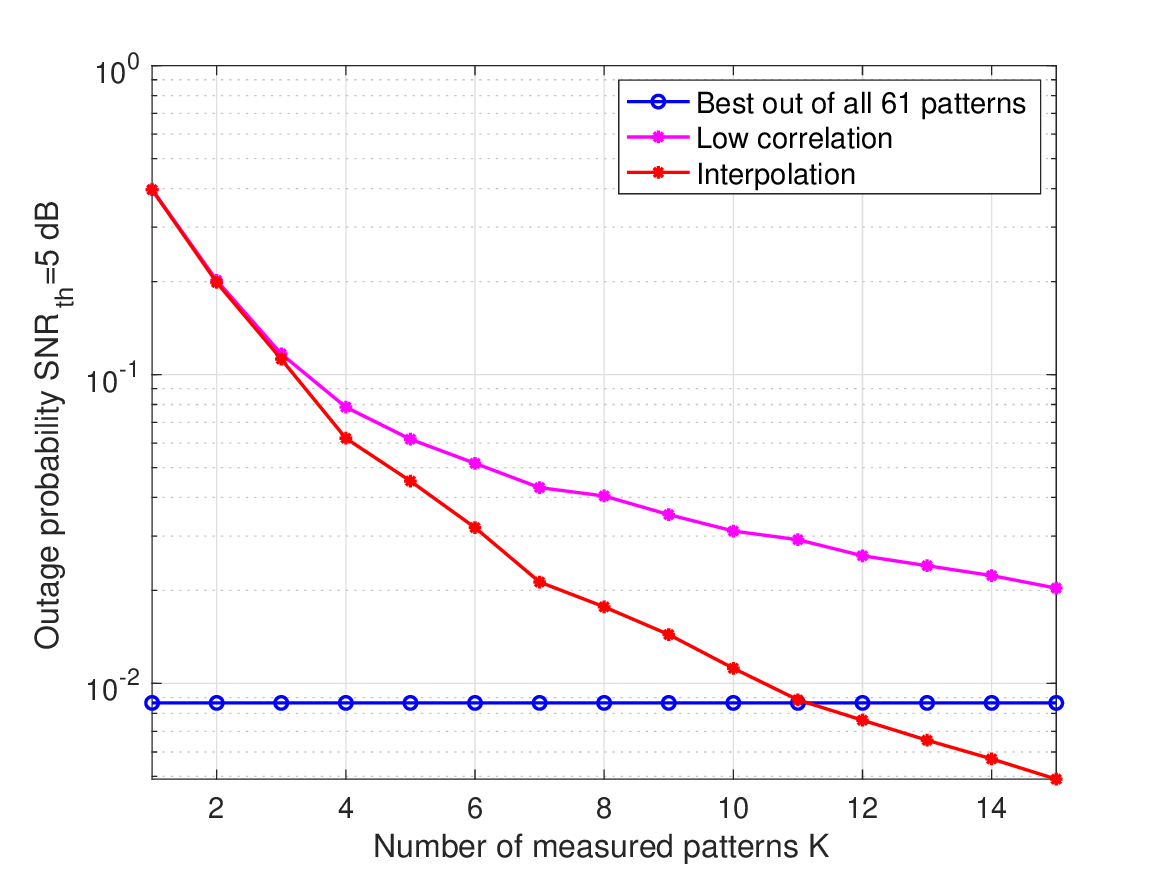}
    \caption{Outage probability as a function of $K$ for $\mathrm{SNR}=8$dB and $\mathrm{SNR}_{\mathrm{th}}=5$dB.}
    \label{fig:outage}
\end{figure}

Figure~\ref{fig:outage} shows the outage probability for an SNR threshold of $\mathrm{SNR}_{\mathrm{th}}=5$~dB at $\mathrm{SNR}=8$~dB as a function of the number of observed radiation patterns $K$. Both low-correlation selection and the proposed interpolation scheme achieve lower outage probabilities as more patterns are observed. Notably, for $K>10$, the proposed interpolation scheme surpasses exhaustive scanning while requiring substantially fewer channel observations and, consequently, lower pilot overhead. 

\begin{figure}[ht]
    \centering
    \includegraphics[width=1.0\linewidth]{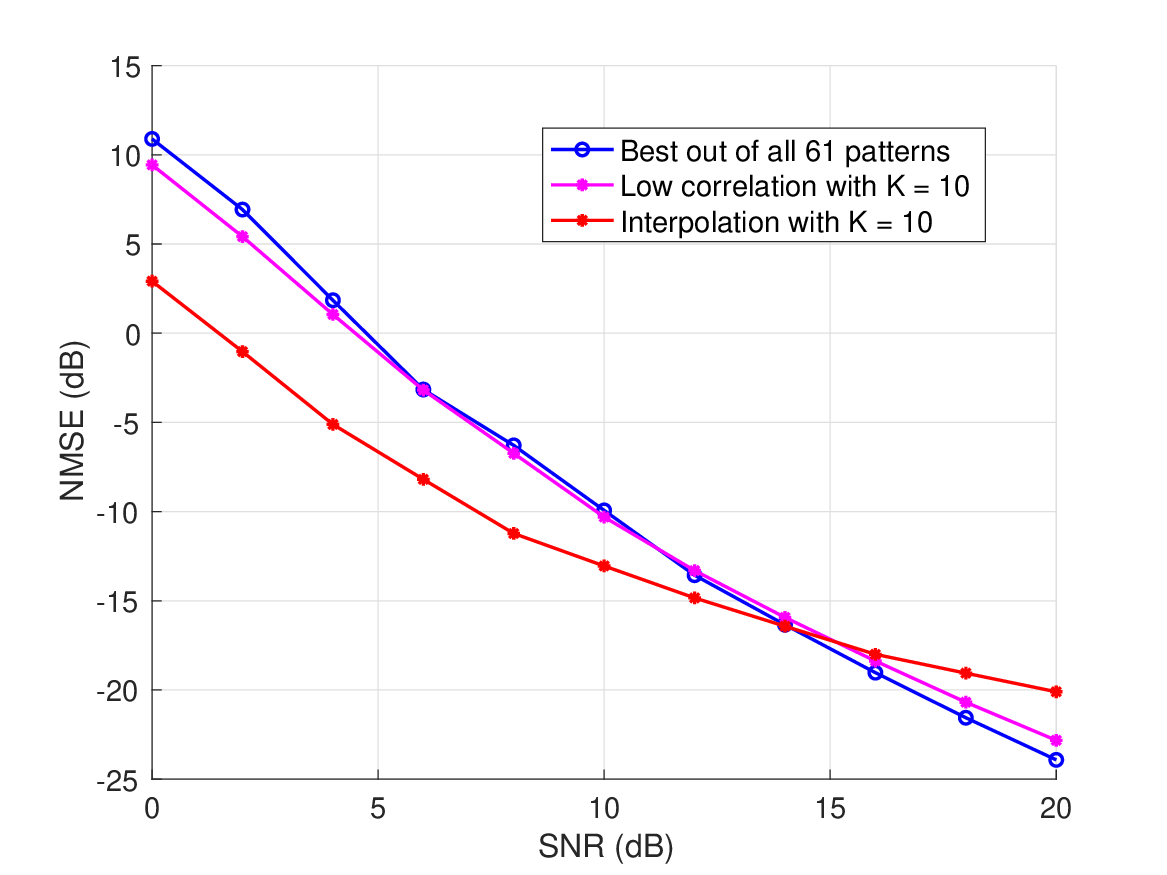}
    \caption{NMSE as a function of SNR for $K=10$.}
    \label{fig:NMSE}
\end{figure}

Finally, Fig.~\ref{fig:NMSE} compares the \gls{NMSE} of the channel estimates obtained using the three selection techniques as a function of \gls{SNR}. The interpolation-based method consistently outperforms the other approaches, with the most significant gains observed at low \gls{SNR}. This improvement can be attributed to the sensitivity of exhaustive scanning to noise-induced peaks. In contrast, the interpolation-based approach exploits the correlation among the channel responses associated with different pixel configurations, enabling more reliable estimation of the unobserved configurations. At very high \gls{SNR}, however, the effect of noise becomes negligible, and interpolation errors become the dominant source of estimation inaccuracy. In this regime, particularly when highly accurate channel estimation is required, increasing the number of observed patterns can help improve the estimation accuracy.

\section{Conclusions}

This paper investigated channel estimation for a beam-steerable reconfigurable pixel antenna, exploiting the correlation between its radiation patterns to reduce channel-observation overhead. A beam-steerable pixel antenna was designed to provide multiple distinct radiation patterns, and their electromagnetic correlation was exploited to interpolate unobserved channel responses from a limited number of measurements. The results demonstrate that accurate pattern selection can be achieved with substantially fewer observations than exhaustive scanning, with improved SNR and NMSE performance, particularly at low SNR.

\section*{Acknowledgment}

\thanks{The research was funded by the Research Council of Finland, Grant number 367569.}

\balance

\vspace{-0.5em}

\bibliographystyle{IEEEtran}
\bibliography{rw_copy}

\end{document}